 \documentclass[sigconf,screen,noacm]{acmart}
\usepackage{amsmath,graphicx,wrapfig,multirow,pbox,makecell,subcaption}

\usepackage[font=small,skip=0pt]{caption}

\AtBeginDocument{%
\providecommand\BibTeX{{%
\normalfont B\kern-0.5em{\scshape i\kern-0.25em b}\kern-0.8em\TeX}}}

\copyrightyear{2020} 
\acmYear{2020} 
\setcopyright{acmcopyright}
\acmConference[BCB '20]{Proceedings of the 11th ACM International Conference on Bioinformatics, Computational Biology and Health Informatics}{September 21--24, 2020}{Virtual Event, USA}
\acmBooktitle{Proceedings of the 11th ACM International Conference on Bioinformatics, Computational Biology and Health Informatics (BCB '20), September 21--24, 2020, Virtual Event, USA}
\acmPrice{15.00}
\acmDOI{10.1145/3388440.3412470}
\acmISBN{978-1-4503-7964-9/20/09}

 \makeatletter
 \def\@copyrightspace{\relax}
 \makeatother

\begin{document}

\title[CNN based segmentation of infarcted regions in acute cerebral stroke patients from CTP imaging]{CNN Based Segmentation of Infarcted Regions in Acute Cerebral Stroke Patients From Computed Tomography Perfusion Imaging}
\author{Luca Tomasetti}
\email{luca.tomasetti@uis.no}
\author{Kjersti Engan}
\email{kjersti.engan@uis.no}
\author{Mahdieh Khanmohammadi}
\email{mahdieh.khanmohammadi@uis.no}
\affiliation{%
\institution{University of Stavanger, Department of Electrical Engineering and Computer Science, BMDLab }
\city{Stavanger}
\state{Norway}
\postcode{4013}
}

\author{Kathinka D{\ae}hli Kurz}
\affiliation{%
\institution{Stavanger University Hospital, Stavanger medical imaging laboratory (SMIL)}
}
\affiliation{%
\institution{University of Stavanger}
\city{Stavanger}
\state{Norway}
\postcode{4013}
}
\email{kathinka.dehli.kurz@sus.no}


\begin{abstract}
More than 13 million people suffer from ischemic cerebral stroke worldwide each year. Thrombolytic treatment can reduce brain damage but has a narrow treatment window. Computed Tomography Perfusion imaging is a commonly used primary assessment tool for stroke patients, and typically the radiologists will evaluate resulting parametric maps to estimate the affected areas, dead tissue (core), and the surrounding tissue at risk (penumbra), to decide further treatments.
Different work has been reported, suggesting thresholds, and semi-automated methods, and in later years deep neural networks, for segmenting infarction areas based on the parametric maps.
However, there is no consensus in terms of which thresholds to use, or how to combine the information from the parametric maps, and the presented methods all have limitations in terms of both accuracy and reproducibility.

We propose a fully automated convolutional neural network based segmentation method that uses the full four-dimensional computed tomography perfusion dataset as input, rather than the pre-filtered parametric maps.
The suggested network is tested on an available dataset as a proof-of-concept, with very encouraging results.
Cross-validated results show averaged Dice score of 0.78 and 0.53, and an area under the receiver operating characteristic curve of 0.97 and 0.94 for penumbra and core respectively.
\end{abstract}

 \settopmatter{printacmref=false}
 \setcopyright{none}
 \renewcommand\footnotetextcopyrightpermission[1]{}
 \pagestyle{plain}

\begin{teaserfigure}
\centering
\includegraphics[width=\textwidth]{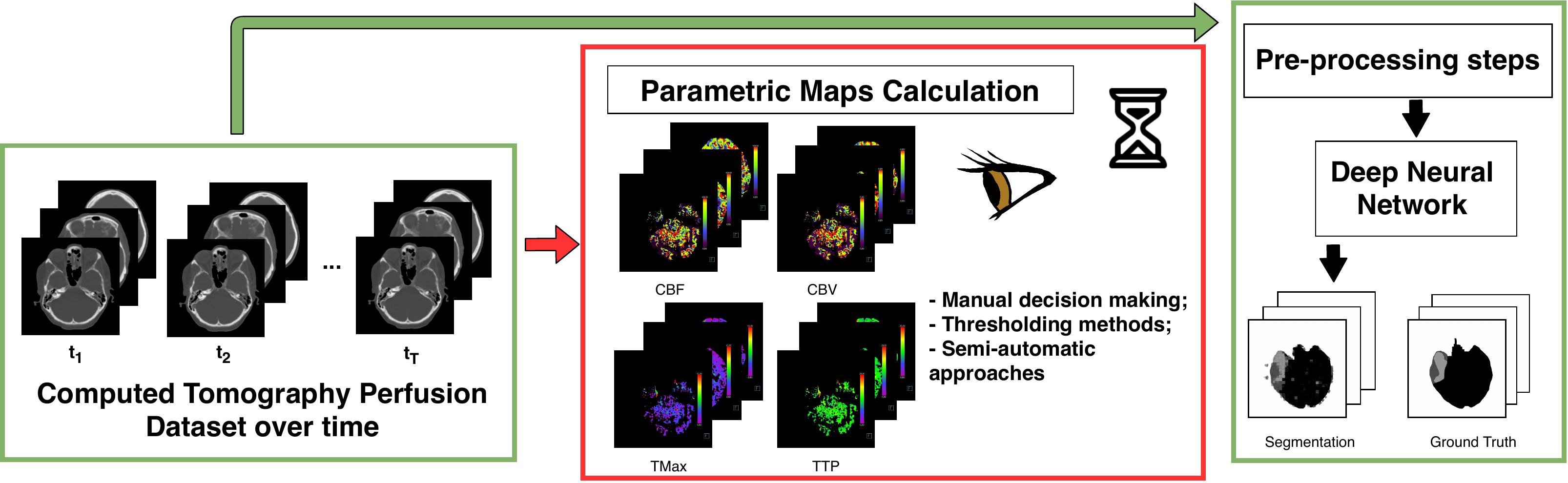}
\caption{A general overview of the various steps involved in the proposed approach (green panels) that bypasses the current methods used by the majority of the radiologists and state-of-the-art approaches based on thresholding and semi-automatic approaches (red panel).}
\label{fig:over}
\end{teaserfigure}

\maketitle
\pagestyle{plain}

\section{Introduction}
\label{sec:intro}

A cerebral stroke is the second most common cause of death among adults worldwide \cite{wang2016global}.
An ischemic stroke can occur if the flow of oxygen-rich blood in an artery to a portion of the brain is occluded.
Ischemic areas are usually heterogeneous with areas that are irreversibly damaged (infarct core) and areas where the tissue is still vital, but critically hypoperfused.
This area is called the penumbra, and is the target for therapy, as restoring blood flow for the penumbra will preserve neurological function for the patient.
This tissue will turn into infarct core if blood flow is not restored timely, and therefore, there is a ``time is brain'' concept \cite{kurz2016radiological}.
A patient can lose up to 1.9 million neurons, 14 billion synapses, and 12 km nerve fibers every minute from the time the stroke happened \cite{saver2006time}; the best treatment window for thrombolysis is estimated to be from 3 up to 4.5 hours from symptom onset \cite{hacke2008thrombolysis}.

Computed Tomography Perfusion (CTP) is a very fast and valuable tool to accurately predict the prognosis of patients in the early stage of the treatment \cite{campbell2013ct}.
Thus, Computed Tomography (CT) is the most commonly used initial imaging method for acute stroke patients.

\begin{figure*}[ht]
\begin{minipage}[b]{\linewidth}
\centering
\centerline{\includegraphics[width=\linewidth]{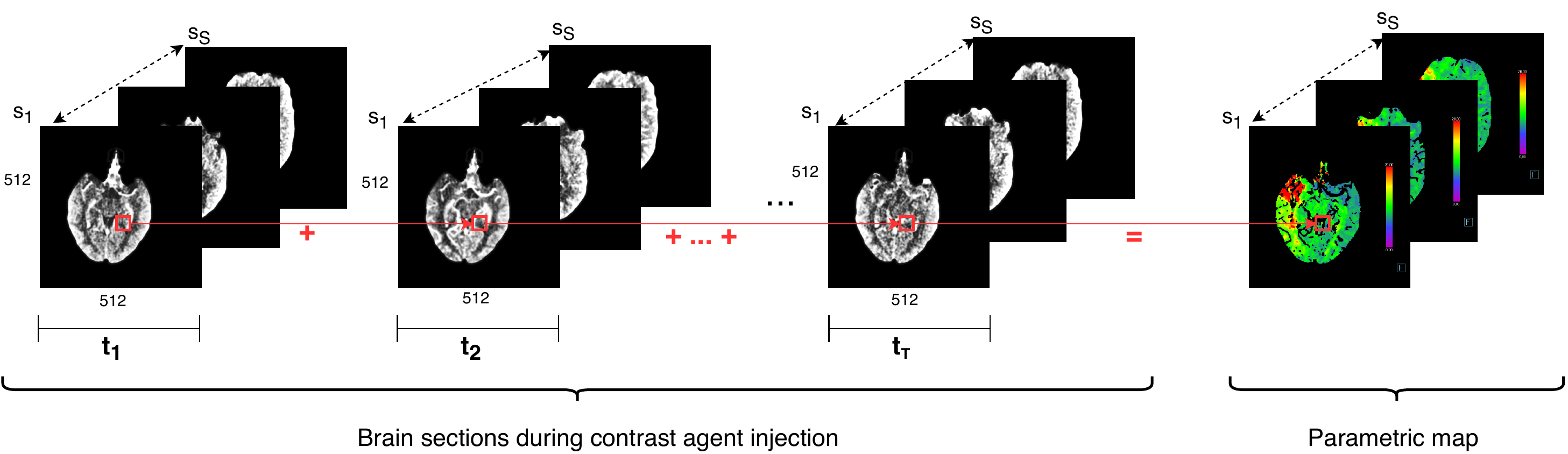}}
\end{minipage}
\caption{The 4D CTP volume is seen as 3D slices through the brain over different time points. To the right, an example of a parametric map (TTP) is seen, calculated from the entire time-span.}
\label{fig:timeseriesbrain}
\end{figure*}

The four-dimensional (4D) examination of CTP is formed by acquiring three-dimensional (3D) CT scans of the brain at many time points during the passage of a contrast agent from the arteries through the capillaries to the veins and then into the venous sinuses \cite{kurz2016radiological, campbell2013ct}.
Parametric color-coded maps are calculated and all the different time points are included in the analyses, describing the changes of blood perfusion and other important factors over time.

Various parametric color-coded maps exist describing precise phenomena in the brain: time to peak (TTP), cerebral blood volume (CBV), relative CBV (rCBV), cerebral blood flow (CBF), relative CBF (rCBF), mean transit time (MTT) and time to maximum $(T_{\text{Max}})$ \cite{kurz2016radiological}.
A meticulous assessment of these parametric maps, performed by a medical specialist, is needed to identify the ischemic regions of a stroke.
These findings will guide the decision on who needs immediate thrombolytic treatment and/or interventional thrombectomy to prevent larger cerebral stroke.
Generation of these parametric maps and their inspections by radiologists to make a final decision about possible patient treatment is time-consuming.

Extensive research has proposed different methods to extract the core and the penumbra based on distinct parametric maps \cite{kasasbeh2019artificial, lucas2018multi, campbell2012comparison, cereda2016benchmarking, ma2019thrombolysis, lin2014comparison, wintermark2006perfusion}.
Campbell et al. \cite{campbell2012comparison}, Cereda et al. \cite{cereda2016benchmarking}, Ma et al. \cite{ma2019thrombolysis}, Lin et al. \cite{lin2014comparison}, and Wintermark et al.\cite{wintermark2006perfusion} are all examples of researches that use extensive and different thresholding approaches to define the infarct regions; hence, a global consensus in the CTP parameters does not exist to identify the infarcted core and penumbra.

In all the mentioned studies, the gold-standard was found from Magnetic Resonance Imaging (MRI) using the Diffusion-weighted Imaging (DWI) sequences generated some hours after the acquisition of CTP images.
DWI sequences highlight the irreversibly ischemic areas; however, the delay in the acquisition of these images may cause a mismatch between what is the infarcted region of the gold-standard, and what was the actual hypoperfused region at the time of the CTP acquisition, which may create an important obstacle in the diagnostic work-up of these patients \cite{cereda2016benchmarking}.
An improved evaluation of these regions may lead to a better selection of patients for treatment in the acute phase.

Image segmentation approaches using convolutional neural networks (CNN) methods, as the backbone, have become a leading research field in the past years.
They have shown to provide promising results in many biomedical imaging applications, such as the famous U-net model \cite{ronneberger2015u} for two-dimensional (2D) segmentation of neuronal structures in electron microscopic stacks, the V-Net structure \cite{milletari2016v} introduced to segment prostate
MRI volumes, the 3D U-Net network for volumetric segmentation by {\c{C}}i{\c{c}}ek et al. \cite{cciccek20163d}, and the 3D CNN architecture for brain lesion segmentation suggested by Kamnitsas et al. \cite{kamnitsas2017efficient}.
Specifically, in the context of stroke patients, Vargas et al. \cite{vargas2019initial} proposed the use of CNN to predict the presence of perfusion deficit; Kasasbeh et al. \cite{kasasbeh2019artificial} used a semi-automatic approach based on a CNN with the parametric maps as input.
Lucas et al. \cite{lucas2018multi} introduced a CNN based on the 2D U-Net with multi-scale information to segment the infarct core inside the brain, also using a set of parametric maps as input and other information.
Also, CNN has been explored on the use of Computed Tomography Angiography (CTA) images, enhancing occlusions in the blood vessels \cite{oman20193d, barman2019determining, sheth2019machine}.

In this work, we propose a fully automatic CNN-based method to segment both the infarct core and penumbra regions, using the complete 4D set of CTP scans over the injection period.
Our method contributes to the following:
\begin{enumerate}
\item It uses CTP slices over time as input since this is normally the first investigation a patient suspected to stroke goes through \cite{kurz2016radiological, european2008guidelines}.
\item The 4D set of CTP scans contains original information compared to the calculated parametric maps.
\item Being independent from parametric maps saves essential time during the investigation of a patient's condition.
\item Our approach is based on a 3D CNN in a bottleneck model inspired by the U-Net \cite{ronneberger2015u} using as gold-standard manually annotated images generated by an experienced radiologist after studying the parametric maps.
\end{enumerate}

To the best of our knowledge, this is the first attempt that uses directly CTP brain slices, and segments automatically the infarct core and penumbra regions together, both essential to guide the decision of the patient's treatment.


\section{Data Material}
\label{sec:methods}

The data material consists of 4230 CTP slices, stored as DICOM files, from 10 pseudonymized patients with large vessel occlusions collected at Stavanger University Hospital between 2014 and 2015.
The mean age was 70.4 years (SD, 8.9), 60\% were female and 50\% of the patients had an acute stroke in the left hemisphere. The median baseline National Institutes of Health Stroke Scale (NIHSS) score was 14.5.

Fig. \ref{fig:timeseriesbrain} presents an overview of a 4D CTP brain; slices of size $(512\times512\times S)$, where $S$ is the number of slices $(s_1, s_2, \ldots, s_S)$, during the acquisition time $t_1, t_2, \ldots, t_T$ after injecting contrast agent in a cubital vein.
Using the information on a sample section over time, from $t_1$ to $t_T$, the parametric maps are generated.

\subsection{Imaging protocol}
CTP images were acquired during injection of 40 ml iodine containing contrast agent (Omnipaque 350 mg/ml) and 40 ml isotonic saline in a cubital vein with a flow rate of 6 ml/s.
The delay for the first slice acquisition was 4 seconds.
A number between 13 and 22 slices $(S)$, covering all brain, is acquired from each patient with a slice thickness of 5 mm.
Each slice was captured 30 times ($t_1, \ldots, t_T$, where $T=30$) over the injection period.
The total number of images per patient is between 390 and 660, depending on the number of slices acquired.
Eight patients were subjected to 416 mA for the x-ray tube current, while two patients were subjected to 350 mA; all of them were exposed to 80 kVp.

\begin{figure}[h]
\begin{minipage}[b]{\linewidth}
\centerline{\includegraphics[width=\linewidth]{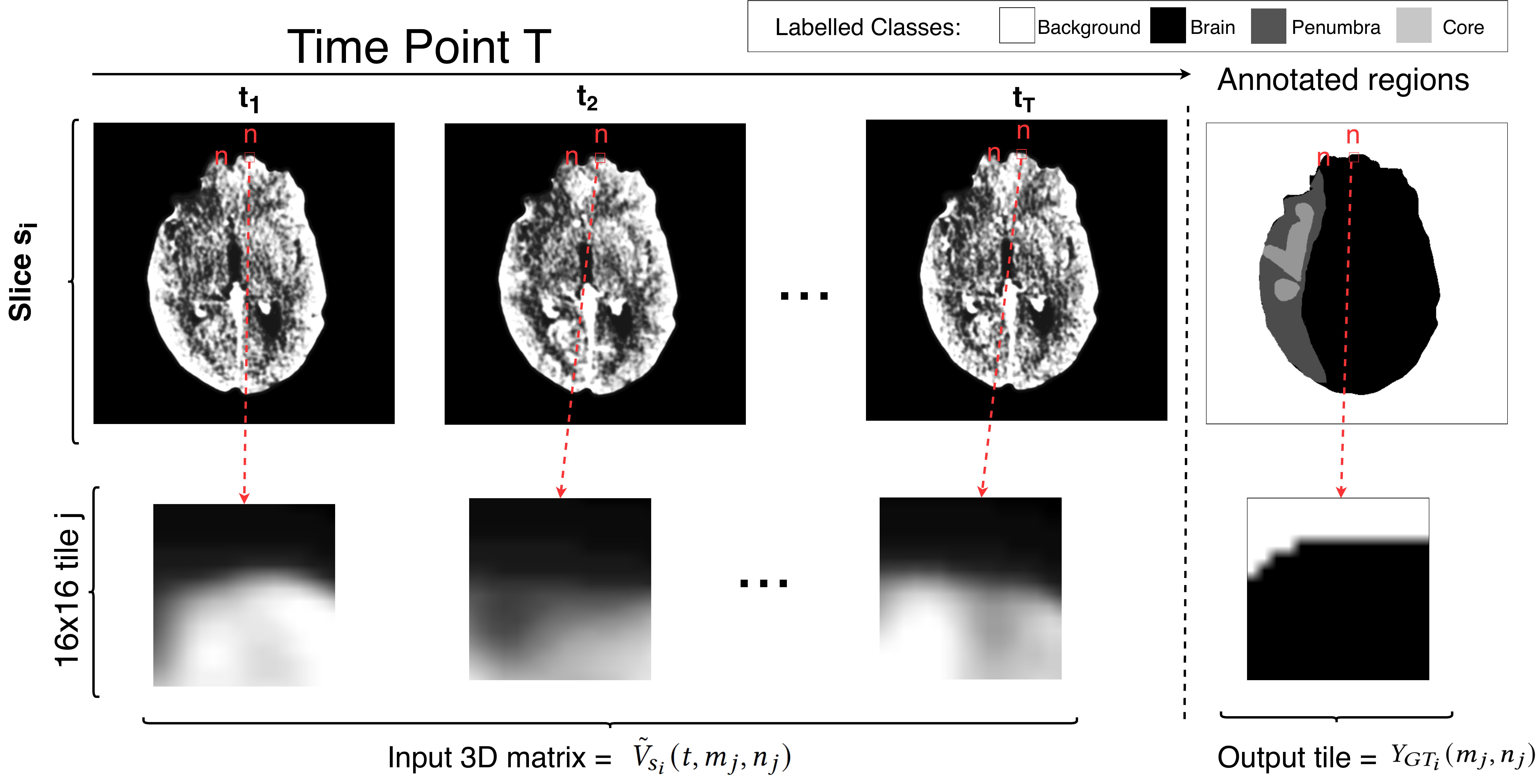}}
\end{minipage}
\caption{A brain slice $s_i$ and the related ground truth image $Y_{GT_i}(m,n)$ are processed through a sliding window technique to create a series of 3D matrices $\tilde{V}_{s_i}(t,m_j,n_j)$ for each tile $j$ and the relative output $Y_{GT_i}(m_j,n_j)$.}
\label{fig:slidwind}
\end{figure}

The ground truth for each brain slice $s_i$ is an image with manually annotated infarcted areas realized by an experienced radiologist from the Stavanger University Hospital after a detailed study of the corresponding parametric maps: TTP, CBV, CBF, and $T_{\text{Max}}$.
The annotated areas in a ground truth image (rightmost column of Fig. \ref{fig:slidwind}) are the brain without stroke symptoms (black), the penumbra (dark gray), the core (light gray) and the background (white), with a target of 0, 76, 150, 255 respectively in grayscale pixel value.

\section{Proposed Method}
\label{sec:met}

We propose a method for fully automatic segmentation of the core area of the stroke as well as the penumbra area from the 4D CTP dataset with a CNN founded method.

Let a CTP data be defined as, $V \in Z^4$, 4D signal of dimension $(T \times M \times N \times S)$.
$T$ is the number of time points $(T=30)$, $M \times N$ is the dimension of each 2D image slice and $S$ the number of slices in the brain.
The image slices are of size $512\times 512$ with a resolution of 0.4258 mm/pixel.
The slice thickness is 5 mm, and as such, the volume has a much lower resolution in the $z$ dimension. At this stage we suggest doing a slice by slice segmentation, but using the full-time series of images of each slice.

Let $V_{s_i}(t,m,n)$ denote the 3D data for slice $s_i$ where $t$ denotes the time index, and $(m,n)$ the spatial image coordinates.
After pre-processing steps (see Sec. \ref{sec:preproc}) it is called $\tilde{V}_{s_i}(t,m,n)$.
We define $A_j$ as a set of image coordinate pairs, $(m,n)$, corresponding to a tile number $j$ of size $16\times 16$ pixel, which extracted from a slice $s_i$. Thus $\tilde{V}_{s_i}(t,m,n)_{(m,n)\in A_{j}}$ gives a volume of size $(30\times 16 \times 16)$ from slice $s_i$ to be used as the input of the neural network.
For notation simplicitiy we write: ${\tilde{V}_{s_i}(t,m,n)}_{(m,n)\in A_j} = \tilde{V}_{s_i}(t,m_j,n_j)$ in the reminder of the paper.
For the 3D output and for the correlated ground truth, the notation becomes $Y_{s_i}(m,n)$ and $Y_{GT_i}(m,n)$ for the entire slice $s_i$; while for tile $j$ the notation becomes $Y_{s_i}(m_j,n_j)$ and $Y_{GT_i}(m_j,n_j)$ respectively.
Note that the time dimension is no longer there, as the data is transformed from 4D to 3D through the analysis.

Due to the limited size of the dataset, a sliding window technique is applied to each slice $s_i$ over the spatial coordinates during training to increase the sample size and create a complete set of 3D matrices composed of $16 \times 16$ tiles.
During prediction using the learning model, tiles can be overlapping, have different strides, or be non-overlapping, this can be taken into account when composing the final output.

An example of one input and the relative ground truth image of the proposed network is given in Fig. \ref{fig:slidwind}: a brain slice $s_i$ is processed through the time series $(t_1, t_2 \ldots, t_{T})$ to generate a 3D matrix $\tilde{V}_{s_i}(t,m_j,n_j)$, after pre-processing steps.
The ground truth, described as $Y_{GT_i}(m_j,n_j)$, represents tile $j$ in the manually annotated brain slice (rightmost column of Fig. \ref{fig:slidwind}).

\subsection{Pre-processing steps}
\label{sec:preproc}
A series of pre-processing steps are done before the CNN module and before applying the overlapping sliding window technique to the dataset.
Firstly $V_{s_i}(t,m,n)$ is pairwise registered over the time sequence $(t_1, \ldots, t_T)$ by a robust similarity-based method \cite{goshtasby20052}.
Subsequently, the skull is removed using a watershed segmentation algorithm \cite{vincent1991watersheds}.
Let $\hat{V}_{s_i}(t,m,n)$ denote the 3D (2D+time) CTP data for slice $s_i$, after all these steps.

At last, $\hat{V}_{s_i}(t,m,n)$ is given as input to a contrast enhancement algorithm, which involves three steps:

\begin{enumerate}
\item image intensity adjustment to increase the contrast by saturating 1\% of the image to low and high intensities;
\item histogram equalization \cite{kim1997contrast} to distribute the intensities;
\item image normalization to harmonize the contrast among all the images.
\end{enumerate}

The input after the pre-processing steps is defined as $\tilde{V}_{s_i}(t,m,n)$.

\subsection{Architectures}
Due to a limited amount of ground truth labeled data to train on, we pre-test, as the first setup, three different CNN-based architectures followed by fully-connected layers for classifier network, where an image block is classified as belonging to the class ``core'', ``penumbra'', ``brain'', or ``background''.
The compressor part of the best architecture for classification is used as the compressor part in a segmentation network, where a decompressor part is added, but producing a 3D output from the 4D input.
All the architectures are implemented and trained with the same preprocessed input; they also produce outputs interpolated to the same size.

\begin{figure}[ht]
\centering
\centerline{\includegraphics[width=.82\linewidth]{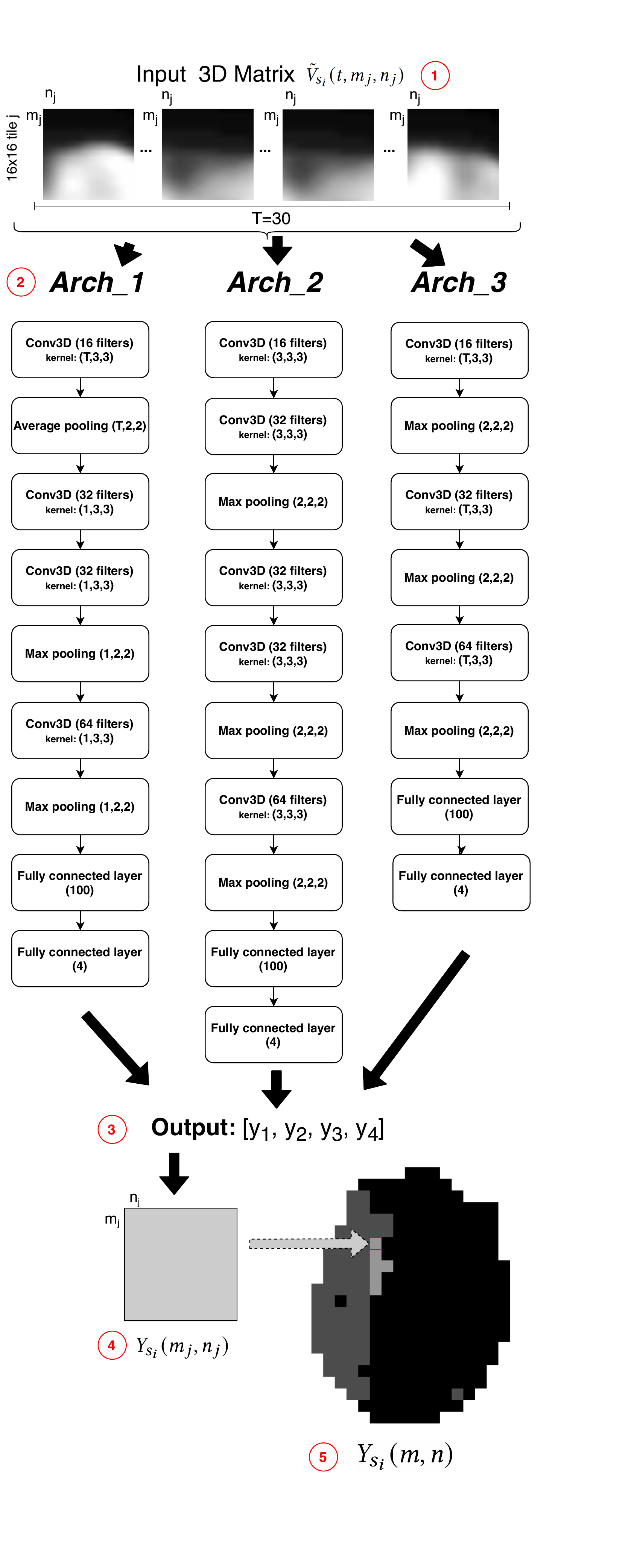}}
\caption{Pipeline overview of the pre-test of different architectures together with the details of their layers. (1) shows the input structure, (2) defines the layers of the three architectures, (3) is the probability vector output, (4) displays the tile expansion for the class with the highest probability value and (5) represents the final image $Y_{s_i}(m,n)$.}
\label{fig:arch3class}
\end{figure}

A description of the various CNN based classifier networks is given in Fig. \ref{fig:arch3class}, including the pipeline involved in the pre-test process.
\textit{Arch\_1}, \textit{2} and \textit{3} are using the same input described in Fig. \ref{fig:slidwind} (marked with (1) in Fig. \ref{fig:arch3class}).
The neural networks generate a probability vector of four elements $[y_1, y_2, y_3, y_4]$ for each input tile $j$ $(T\times16\times16)$ (3).
The highest prediction value is selected and expanded in the $16\times16$ tile $j$ with the corresponding labeled grayscale value $Y_{s_i}(m_j,n_j)$ (4).
The final outcome is a composed pixelated image $Y_{s_i}(m,n)$ of all the generated $16\times16$ tiles (5).

\begin{table}[htb]
\caption{Comparison of the three classifiers based on the average accuracy, precision, recall, and computational time for each tested patient.}
\label{tab:3class}
\centering
\resizebox{\linewidth}{!}{%
\begin{tabular}{cccccc}
\toprule
\textbf{Architecture} & \textbf{Acc.(\%)} & \textbf{Prec. (\%)} & \textbf{Recall (\%)} & \textbf{F1 Score (\%)} & \textbf{Time (s)} \\
\midrule
\textit{Arch\_1} & \textbf{92.34 $\pm$ 4.35} & \textbf{83.6} & \textbf{75.1} & \textbf{79.12} & $\sim$ \textbf{600} \\ \hline
\textit{Arch\_2} & 91.29 $\pm$ 4.71 & 82.4 & 73.8 & 77.86 & $\sim$ \textbf{600} \\ \hline
\textit{Arch\_3} & 89.67 $\pm$ 4.44 & 77.1 & 70.4 & 73.60 & $\sim700$ \\
\bottomrule
\end{tabular}}
\end{table}

\begin{table}[h!]
\caption{Comparison of the three pre-tested architectures and the implemented network (\textit{mJ-Net}): \textit{conv3D} represents a 3D convolution layer, \textit{max\_pool} and \textit{avg\_pool} are a max and an average pooling layer respectively, \textit{fully\_conn} is the fully connected layer and \textit{concat \& transpose} is the upsamples layer. $T$ is the number of time points in the 3D matrix (30).}
\label{tab:diffarch}
\resizebox{\linewidth}{!}{%
\begin{tabular}{lcccc}
\toprule
\multicolumn{1}{c}{} & \textbf{Arch\_1} & \textbf{Arch\_2} & \textbf{Arch\_3} & \textbf{mJ-Net} \\
\midrule
\textbf{Input} & 3D matrices & 3D matrices & 3D matrices & 3D matrices \\ \hline
\textbf{Output} & {\begin{tabular}[c]{@{}l@{}} Probability \\ vector 4$\times$1 \end{tabular}} & {\begin{tabular}[c]{@{}l@{}} Probability \\ vector 4$\times$1 \end{tabular}} & {\begin{tabular}[c]{@{}l@{}} Probability \\ vector 4$\times$1 \end{tabular}} & {\begin{tabular}[c]{@{}l@{}} $16\times16$ \\ labelled tile \end{tabular}} \\ \hline
\textbf{\begin{tabular}[c]{@{}l@{}} \# classes \\ per tile\end{tabular}} & 1 & 1 & 1 & $\ge$ 1 \\ \hline
\textbf{\begin{tabular}[c]{@{}l@{}}Problem\\ Approach\end{tabular}} & Classification & Classification & Classification & Segmentation \\ \hline
\textbf{\# Layers} & 9 & 10 & 8 & 19 \\ \hline
\textbf{\# Parameters} & 203,320 & 773,384 & 63,312 & 981,553 \\ \hline
\textbf{Layers} & {\begin{tabular}[c]{@{}l@{}} conv3D + \\ avg\_pool + \\ (2 $\times$ conv3D) + \\ max\_pool + \\ conv3D + \\ max\_pool + \\ (2 $\times$ fully\_conn) \end{tabular}} & {\begin{tabular}[c]{@{}l@{}} 2 $\times$ [ (2 $\times$ conv3D) + \\ max\_pool ] + \\ conv3D + \\ max\_pool + \\ (2 $\times$ fully\_conn) \end{tabular}} & {\begin{tabular}[c]{@{}l@{}} $3 \times$ (conv3D + \\ max\_pool) + \\ (2 $\times$ fully\_conn) \end{tabular}} & {\begin{tabular}[c]{@{}l@{}} conv3D + \\ avg\_pool + \\ 2 $\times$ (2 $\times$ conv3D + \\ max\_pool) + \\ 2 $\times$ conv3D + \\ 2 $\times$ [ (concat \& transpose) + \\ (2 $\times$ conv3D) + \\ max\_pool ] + \\ conv3D \end{tabular}} \\ \hline
\textbf{Kernel Size} & {\begin{tabular}[c]{@{}l@{}} (T,3,3) first layer \\ (1,3,3) other layers \end{tabular}} & (3,3,3) & (T,3,3) & {\begin{tabular}[c]{@{}l@{}} (T,3,3) first layer \\ (1,3,3) other layers \\ (1,2,2) concat \& transpose layers \end{tabular}} \\ \hline
\textbf{\begin{tabular}[c]{@{}l@{}}Pooling\\ Window\end{tabular}} & {\begin{tabular}[c]{@{}l@{}} (T,2,2) second layer \\ (1,2,2) other layers \end{tabular}} & (2,2,2) & (2,2,2) & {\begin{tabular}[c]{@{}l@{}} (T,1,1) avg\_pool \\ (1,2,2) first two max\_pool \\ (2,1,1) last two max\_pool \end{tabular}} \\ \hline
\textbf{\begin{tabular}[c]{@{}l@{}}\# of filters \\per \textit{conv3D} \end{tabular}} & {\begin{tabular}[c]{@{}l@{}} 16, 32, 32, 64 \end{tabular}} & {\begin{tabular}[c]{@{}l@{}} 16, 32, 32, 32, 64 \end{tabular}}\ & {\begin{tabular}[c]{@{}l@{}} 16, 32, 64 \end{tabular}}\ & {\begin{tabular}[c]{@{}l@{}} 16, 32, 64, 64, 128, \\128, 256, 128, 64, 32, 32 \end{tabular}} \\
\bottomrule
\end{tabular}%
}
\end{table}

The various architectures are illustrated in Fig. \ref{fig:arch3class}.
\textit{Arch\_1} reduces the third dimension of its input into a 2D matrix after the \textit{avg\_pool} layer to average the entire time dimension of the 3D tile into a 2D matrix.
The implementation of \textit{Arch\_1} was mainly done to decrease the learning parameters in each layer for reducing computational time while at the same time have useful information for training purposes.
On the contrary, \textit{Arch\_2} and \textit{3} work always with a 3D matrix in their layers.
\textit{Arch\_2} uses a kernel size in each convolution layer of $(3,3,3)$ while \textit{Arch\_3} adopts a kernel size of $(T,3,3)$; thus, in each convolutional layer, \textit{Arch\_3} convolves a higher number of time points $t$ in the $\tilde{V}_{s_i}(t,m_j,n_j)$ matrix than \textit{Arch\_2}.

Our final objective is to find a fast and automatic segmentation approach.
Following this motivation, the classifier model that yields the best results (\textit{Arch\_1}, as displayed in Table \ref{tab:3class}) was further expanded to a segmentation model, inspired both by the U-Net \cite{ronneberger2015u} and the V-Net \cite{milletari2016v}.
We call it ``mirror J-Net'' (\textit{mJ-Net}) because of its particular shape.
Table \ref{tab:diffarch} presents a comparison of the three classifier structures and the derived \textit{mJ-Net} architecture, showing in detail the design of all the layers involved in each model.

\begin{figure}[ht]
\centering
\centerline{\includegraphics[width=.97\linewidth]{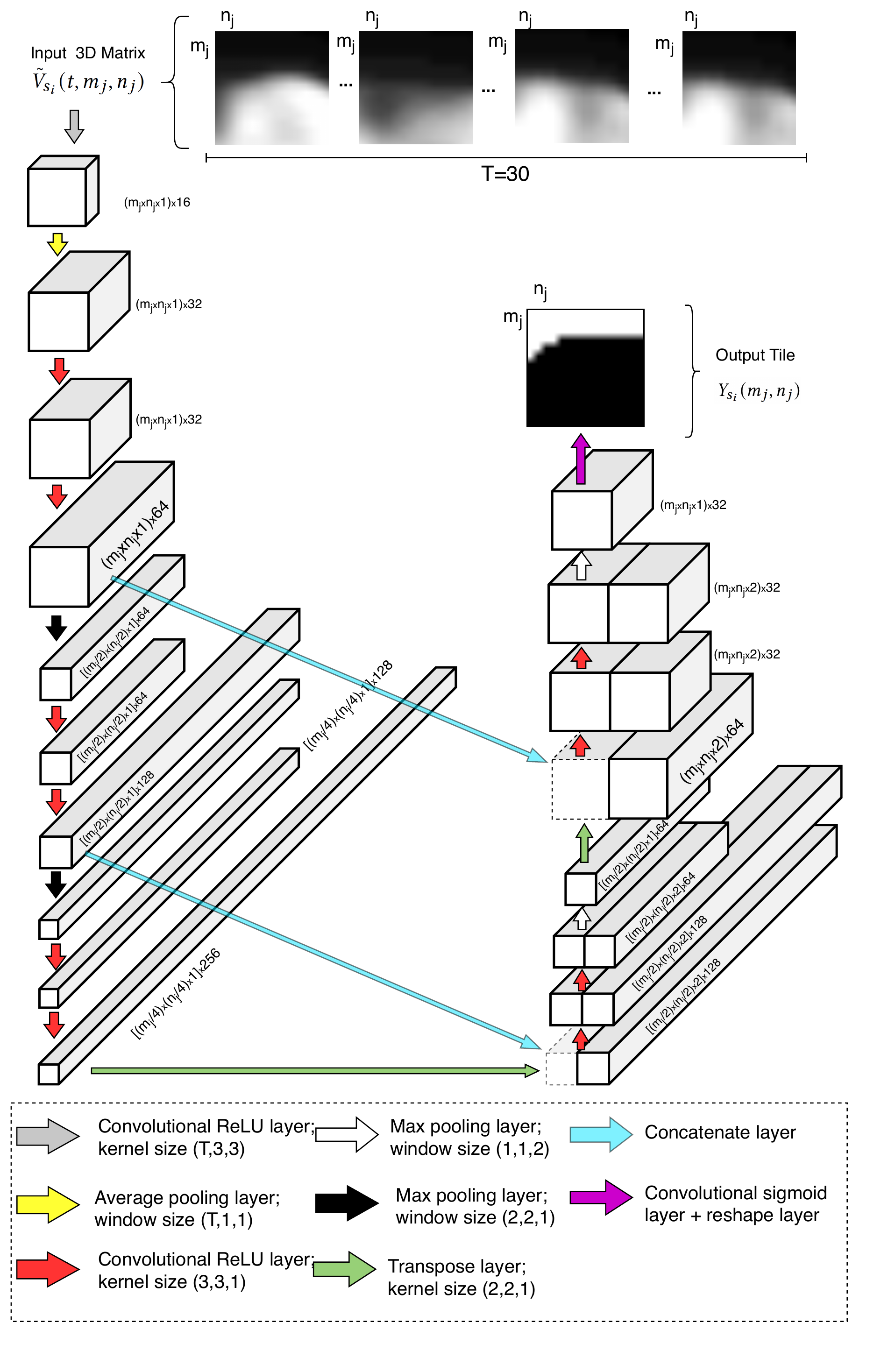}}
\caption{Detailed overview of the mirror J-Net (\textit{mJ-Net}).}
\label{fig:overdetail}
\end{figure}

\begin{figure*}[ht]
\begin{subfigure}[b]{0.5\linewidth}
\begin{subfigure}[b]{\linewidth}
\includegraphics[width=\linewidth]{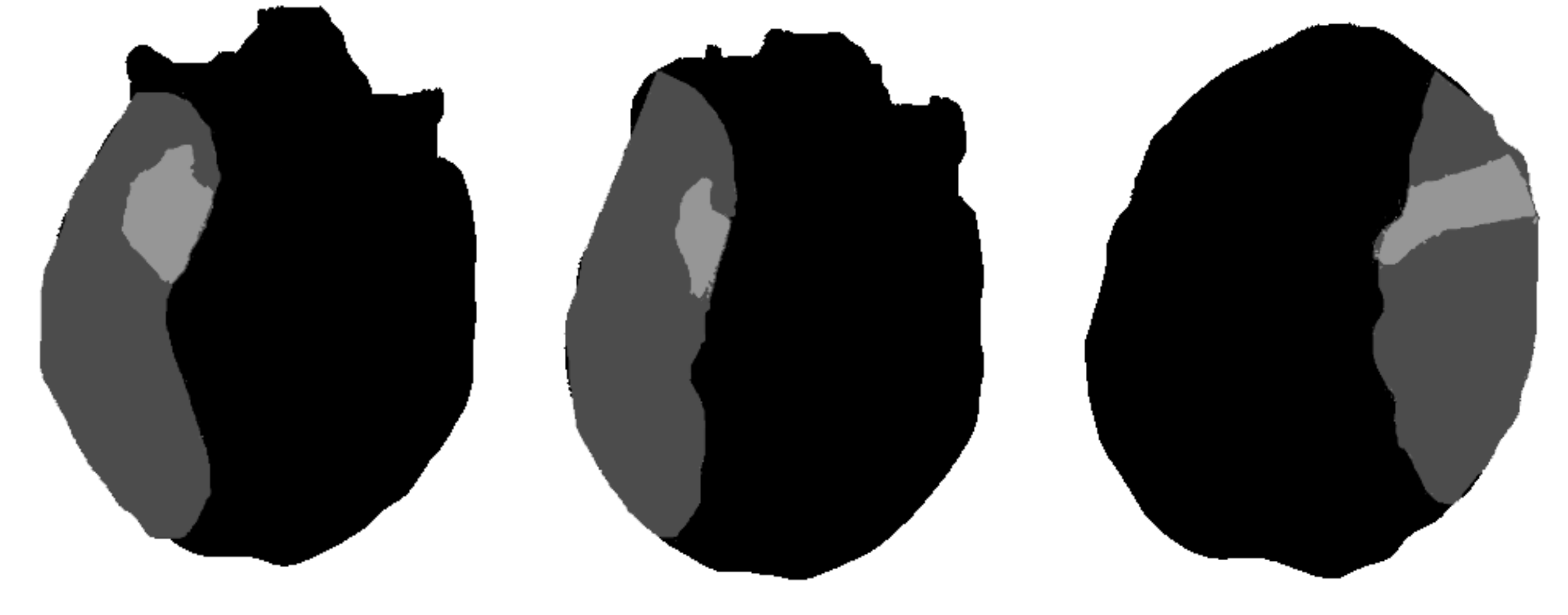}
\caption{ Manual annotated image}
\end{subfigure}
\begin{subfigure}[b]{\linewidth}
\includegraphics[width=\linewidth]{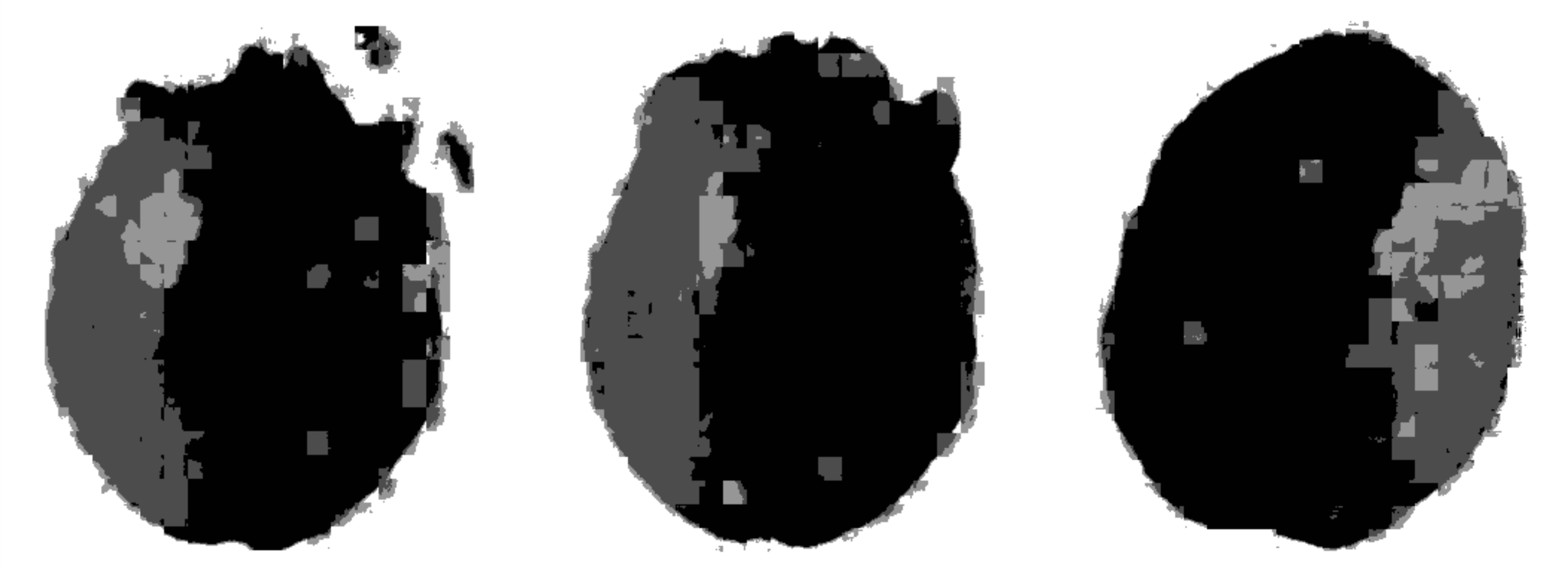}
\caption{Automatically predicted images}
\end{subfigure}
\end{subfigure}
\begin{subfigure}[b]{0.45\linewidth}
\includegraphics[width=\linewidth]{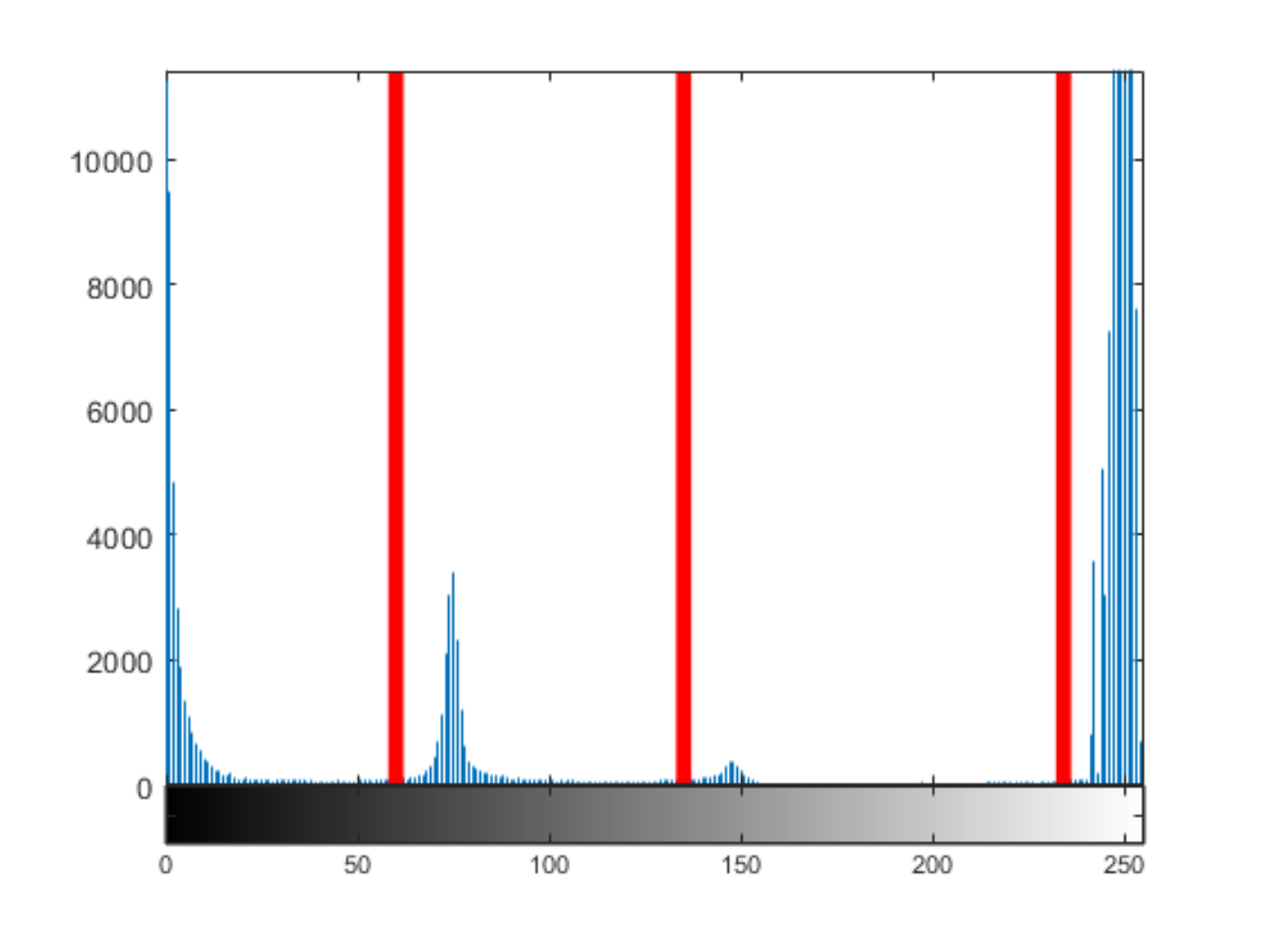}
\caption{Histogram of one predicted image.}
\end{subfigure}

\caption{Example of three brain slices $s_i$ comparison between (a) their ground truth, and (b) the deep neural network output using the best \textit{mJ-Net} model; (c) is the histogram used to divide the various classes.}
\label{fig:expres}
\end{figure*}

A detailed view of the \textit{mJ-Net} model is given in Fig. \ref{fig:overdetail}.
The output from \textit{mJ-Net} is a matrix of dimension $16\times16$.
Each value in the matrix corresponds to a probability number which is multiplied by $255$ to represent a grayscale pixel.
A post-processing step is performed to combine all outputs and generate the final segmented image, which is represented as a grayscale labeled image with the same dimension as the initial CTP images ($512\times512$ pixels) and the labels ``core'', ``penumbra'', ``brain'', ``background'' (Fig. \ref{fig:expres}(b)).

The \textit{mJ-Net} network is formed by a convolutional and a de-convolutional part.
The convolutional part (left side on the Fig. \ref{fig:overdetail}) presents a very similar structure as the \textit{Arch\_1}: all the layers are the same with the only exception of the last fully-connected layers, which are substituted by a new convolutional layer useful for the second part of the network; moreover, the average pooling operation in the second layer flats the time dimension of the input into a 2D matrix in the same way as the \textit{Arch\_1}.
The de-convolutional part (right side in Fig. \ref{fig:overdetail}) is composed of concatenations of a transpose layer and the output of one of the previous convolution layers with the same dimension; convolutional layers and max-pooling operations are performed after the creation of these layers to increase the resolution of the output $Y_{s_i}(m_j,n_j)$.
The structure of the \textit{mJ-Net} is inspired by the U-Net and the V-Net architectures, and has some similarities with these structures and their layers.
Nevertheless, while the U-Net and the V-Net preserve the dimension of the input and output (respectively 2D for the U-Net and 3D for the V-Net), the \textit{mJ-Net} takes a 3D volume as input and produces a 2D image as output.

\section{Experiments and Results}
\label{sec:results}

Due to the limited amount of training data, a leave-one-patient-out cross-validation is performed.
Also, the images are tiled into $(T\times16\times16)$ regions with an overlapping sliding window technique.
The ``core'' class is the most important infarcted region to detect, and the number of tiles in this region is a very small portion of the entire dataset.
Therefore a series of data augmentation techniques were implemented to extend the number of core class tiles and improve the balance between the classes during training.
These techniques consist of rotation or mirroring of the time-series tiles only if the corresponding ground truth section is classified as a ``core'' area.
The total number of volumes $\tilde{V}_{s_i}(t,m_j,n_j)$ generated after augmentation is 1,086,030, divided between the following classes: 28.3\% background, 54.9\% brain, 11.7\% penumbra and 5.1\% core.

During the learning of the networks, two different optimizer functions were tested: the Adaptive moment estimation (Adam) \cite{kingma2014adam}, and the stochastic gradient descent (SGD) optimizer function.
The best results were achieved with the SGD optimizer using a learning rate sets to 0.01 and the Nesterov momentum sets to 0.9 \cite{nesterov2013introductory}.
The statistical results are shown in Table \ref{tab:compmj}.

All experiments are performed with a GPU Tesla V100-PCIE (32GB) and 100 epochs per training with early stopping if the loss does not decrease for a fixed number of epochs (10).
Keras 2.3.1 was used for the implementation detail of the networks, with Tensorflow 1.14 as backend.
For all the models, leave-one-patient-out cross-validation was performed, using 9 patients (including $\approx$4000 images) for training, and testing on the last patient (about 360-660 images) in each fold.
The training of the \textit{mJ-Net}, for each fold in the leave-one-patient-out cross-validation, needs approximately 22 hours on average to complete. Less than 20 seconds were necessary, on average, to generate a brain slice image for a testing patient.

We choose a modified version of the Dice coefficient metric for the loss function, called the soft Dice coefficient.
This formulation was proven to generate better outcomes for a dataset with unbalanced classes without the need to assign weights to samples to balance the dataset \cite{milletari2016v}.
For each predicted tile $j$, the loss function is defined as:

$$
\text{loss}_j = 1 - \frac{2 \sum_{\forall (m_j,n_j)} | Y_{GT_i}(m_j,n_j) Y_{s_i}(m_j,n_j) | + \epsilon}{ \sum_{\forall (m_j,n_j)} (Y_{GT_i}(m_j,n_j))^2 + \sum_{\forall (m_j,n_j)} (Y_{s_i}(m_j,n_j))^2 + \epsilon}
$$
where $\epsilon>0$ is a small number for numerical stability.
Subsequentially, the cost function is represented as:

$$
\text{cost}_{\text{batch}} = \sum_{j \in \text{batch}} \text{loss}_j
$$

The best classifier architectures, among the three that were pre-tested, was \textit{Arch\_1}, as shown in Table \ref{tab:3class}.
\textit{Arch\_1} yield the best accuracy, precision, F1 score, and recall on average; it also presented a considerable fast computational time for each training epoch.

Fig. \ref{fig:expres}(a) shows three ground truth examples of the manual annotated brain slices $s_i$.
The predicted outcomes of the same brain slices $s_i$ with the best \textit{mJ-Net} after post-processing steps are shown in Fig. \ref{fig:expres}(b).
At the bottom of Fig. \ref{fig:arch3class} (marked with 5), an example of the generated brain slice $s_i$ with \textit{Arch\_1}, after the union of the various tiles; the ground truth image is displayed on the left part of Fig. \ref{fig:expres}(a).

\begin{table*}[ht]
\caption{Comparison of the \textit{mJ-Net} results with literature methods. The input column shows specific parametric maps with the corresponding threshold used as input to extrapolate the infarcted areas, except our approach which uses in input the entire 4D CTP images. \\
*Kasasbeh' results were generetad not with our dataset, however a comparison is still applicable since the NIHSS score of both dataset is almost identical.}
\label{tab:compmj}
\centering
\resizebox{.9\linewidth}{!}{%
\begin{tabular}{cccccccccc}
\toprule
\textbf{Method} & \textbf{Patients} & \textbf{\makecell{Infarcted \\ Area }} & \textbf{\makecell{Dice \\ coef.}} & \textbf{Sens.} & \textbf{Spec.} & \textbf{Prec.} & \textbf{Acc.} & \textbf{AUC} & \textbf{Input} \\ \midrule

\multirow{2}{*}{ mJ-Net (SGD) } & \multirow{2}{*}{10} & Penumbra & \textbf{0.78} & \textbf{0.86} & 0.93 & \textbf{0.72} & \textbf{0.95} & \textbf{0.97} & \multirow{2}{*}{\makecell{ 4D CTP }} \\ 
& & Core & \textbf{0.53} & 0.75 & \textbf{0.98} & \textbf{0.41} & \textbf{0.99} & \textbf{0.94} & \\ \hline

Wintermark et al. \cite{wintermark2006perfusion} & \multirow{1}{*}{10} & Core & 0.14 & 0.72 & 0.90 & 0.08 & 0.90 & 0.86 & CBV$<$33\% \\

\multirow{2}{*}{ Campbell et al. \cite{campbell2012comparison} } & \multirow{2}{*}{10} & Penumbra & 0.40 & 0.30 & 0.96 & 0.62 & 0.85 & 0.71 & $T_{\text{Max}}>$6s\\
& & Core & 0.05 & 0.35 & 0.86 & 0.03 & 0.85 & 0.30 & CBF$<$31\% \& TTP$>$4s \\

\multirow{2}{*}{ Cereda et al. \cite{cereda2016benchmarking} } & \multirow{2}{*}{10} & Penumbra & 0.34 & 0.32 & 0.88 & 0.35 & 0.79 & 0.69 & $T_{\text{Max}}>$4s\\
& & Core & 0.07 & 0.85 & 0.74 & 0.04 & 0.74 & 0.87 & CBF$<$38\% \\

\multirow{2}{*}{ \makecell{Ma et al. \cite{ma2019thrombolysis}, \\ Lin et al. \cite{lin2014comparison}} } & \multirow{2}{*}{10} & Penumbra & 0.37 & 0.26 & \textbf{0.97} & 0.65 & 0.85 & 0.69 & $T_{\text{Max}}>$6s\\
& & Core & 0.09 & 0.80 & 0.81 & 0.05 & 0.80 & 0.87 & CBF$<$30\% \\ \hline

\multirow{1}{*}{Kasasbeh \cite{kasasbeh2019artificial} *} & 128 & Core & 0.48 & \textbf{0.91} & 0.65 & N.A. & N.A. & 0.87 & \makecell{ CBF, CBV, \\ TTP, $T_{\text{Max}}$ \\ \& clinical data} \\ \bottomrule

\end{tabular}
}
\end{table*}

We excluded ``background'' predictions from the calculation of the statistical results to balance the results only with the classes inside the skull.
Since the pixel values are in a [0, 255] domain, we decided to categorize them into four classes.
Based on the histogram of one output image (Fig. \ref{fig:expres}(c)) we choose for the ``brain'' class the domain $[0, 60)$, $[60, 135)$ for ``penumbra'', $[135, 234)$ for ``core'' and the remaining values are ``background''.
We noticed from the histogram that most output values are centered around the target values $\pm 15$, so the threshold is not very sensitive to small variations outside these limits: future work is needed to find optimal thresholds on the training dataset.

The statistical results for the best \textit{mJ-Net} architectures are compared with other reported methods in the literature, as showed in Table \ref{tab:compmj}.
All the thresholding methods \cite{campbell2012comparison, cereda2016benchmarking, ma2019thrombolysis, lin2014comparison, wintermark2006perfusion} were tested on our dataset, excluding the ``background'' from the calculation, to have comparable results.
We also compare our results with the reported results of the recent CNN-based method by Kasasbeh et al. \cite{kasasbeh2019artificial}, however, we could not test it on our dataset because there are no publicly available weights for the proposed neural network.
Nevertheless, since the median NIHSS score was 15 for the 128 patients analyzed in \cite{kasasbeh2019artificial}, and this is approximately the same as the NIHSS score of our dataset, a comparison is still applicable.

\begin{figure}[h!]
\centering
\begin{minipage}[b]{\linewidth}
\centering
\centerline{\includegraphics[width=.92\linewidth]{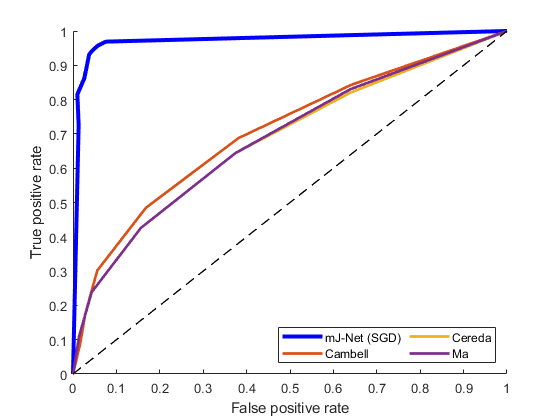}}
\centerline{(a)AUC for penumbra region}\medskip
\end{minipage}
\begin{minipage}[b]{\linewidth}
\centering
\centerline{\includegraphics[width=.92\linewidth]{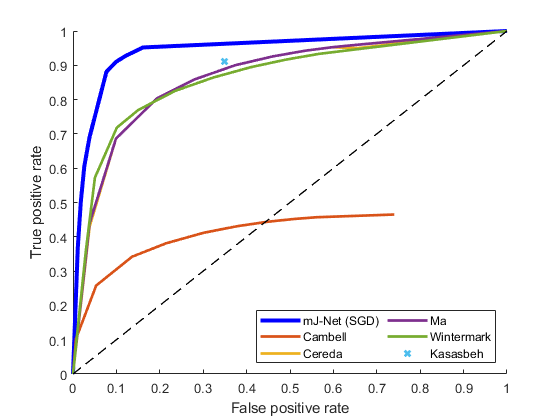}}
\centerline{(b)AUC for core region}\medskip
\end{minipage}
\caption{Comparison of AUC both for (a) penumbra and (b) core regions between the literature methods and the \textit{mJ-Net} model.}
\label{fig:roccurves}
\end{figure}

The optimal \textit{mJ-Net}, based on the SGD optimizer function, produces a Dice coefficient score of 0.78 and 0.53, for penumbra and core respectively.
The area under the receiver operating characteristic curve (AUC), calculated with an expanding upper and lower bound starting from the target value of each class, is 0.97 and 0.94, corresponding to penumbra and core.
A comparison with the AUC of the various reference methods is given in Fig. \ref{fig:roccurves}(a) for the penumbra area and in Fig. \ref{fig:roccurves}(b) for the core region.
Our proposed method, with 4D CTP input, preprocessing and \textit{mJ-Net} for multiclass segmentation, achieved the best results both for penumbra and core.

\section{Discussion}
\label{sec:discussion}

This study aimed to investigate the possibility of detecting the infarcted regions of an ischemic stroke based solely on the 4D CTP data using the proposed neural network.
The performed experiments proved the validity of this network compared with both the referred methods that use thresholding approaches \cite{wintermark2006perfusion,campbell2012comparison, cereda2016benchmarking, ma2019thrombolysis, lin2014comparison} and the semi-automatic method that uses the parametric maps as input \cite{kasasbeh2019artificial}.
The results are promising for both infarcted core and penumbra regions; this solution provides an alternative way for detecting these regions without using thresholding values and it could be used by medical doctors as a support instrument together with the parametric maps.

To our knowledge, this is the first study that tries to segment penumbra and core using the 4D set of CTP scans.
The results achieved by the \textit{mJ-Net} provide the foundations for this objective but further research is necessary to validate the results of the network.

The bottom part of Fig. \ref{fig:arch3class} (marked with (5)) presents one of the outputs of the \textit{Arch\_1}: the areas inside the brain are classified in a pixelated way; this is not helpful for medical decisions during the treatment of a patient.
For this reason, \textit{mJ-Net} was implemented to improve \textit{Arch\_1} to overcome the approximation in the regions' segmentation.
\textit{Arch\_1} was the model that yields the best results among the three classifiers; it also had the best ratio between the predictions and computational time.
The other two architectures presented worse outcomes in comparison with \textit{Arch\_1}; moreover, they have some disadvantages.
\textit{Arch\_2} did not convolve the entire third dimension (time) with its kernel, thus its results were not as accurate as of the other two classifiers.
The number of parameters involved in \textit{Arch\_3} is almost four times larger than the one used in \textit{Arch\_2}: this increases significantly the computational time of \textit{Arch\_3} with improvements in the results of some patients but not significantly close to \textit{Arch\_1} (see Table \ref{tab:3class}).

Our proposed best network displays better results in the Dice coefficient, precision, accuracy, and AUC (Table \ref{tab:compmj}) both for penumbra and core compared with the methods analyzed.
Also, the \textit{mJ-Net} shows high outcomes for sensitivity (``penumbra'') and specificity (``core'') even if these are not the best results for both the infarcted regions, since Kasasbeh et al. \cite{kasasbeh2019artificial} presented a better outcome in the sensitivity for the ``core'' region, while the thresholding values both proposed by Ma et al. \cite{ma2019thrombolysis} and Lin et al. \cite{lin2014comparison} achieved a better specificity for the ``penumbra'' area.

The dataset presents some limitations.
First, due to the small number of patients involved; however, the leave-one-patient-out cross-validation performed over the networks helped to achieve consistent results.
Second, the manual annotations of the ischemic regions are not necessarily providing a perfect ground truth.
The annotations are enclosing the most important regions of the infarcted areas but might leave out small regions of core spread in the penumbra, etc.
However, the annotations are made by a medical expert in the field and reflect how the clinicians are working in a true setting.

\section{Conclusion \& Future Work}
\label{sec:conc}
To our knowledge, this is the first study to use the entire 4D set of CTP brain slices over the injection period to train a 3D CNN to segment the infarcted regions, both core and penumbra, in patients affected by an acute ischemic stroke in a fully automated method.

The present study serves as a proof-of-concept, showing the possibilities of this technology, and it displays promising results with the cross-validated result: average Dice coefficient score of 0.78 and 0.53, and an AUC of 0.97 and 0.94 for penumbra and core respectively.
It shows substantially improved results compared to the different methods combining thresholding of the different parametric maps.
Further research with a larger dataset is required to assert the validity of the proposed architecture and the achieved results.
Another possible future work might be a focus on the evaluation of a more precise ground truth based on all the parametric maps generated during the CTP in combination with medical expert input.
Furthermore, another interesting study for the future should be the test of the proposed method with a dataset composed of patients affected by small or large vessel occlusions to understand how the architecture can segment the infarct regions in a more general dataset.


\bibliographystyle{ACM-Reference-Format}
\bibliography{strings}

\end{document}